\documentclass[twoside]{ilcws08}
\usepackage[latin1]{inputenc}
\usepackage[dvips]{graphicx,epsfig,color}
\usepackage{wrapfig,rotating}
\usepackage{amssymb,amsmath,array}

\begin{document}
\title{
%%%%   Paper title goes here  %%%%%%%%%%%%%%
% Note that nouns are capitalized in the title.
PFA Performance for SiD} %% 
%***********************************************************************
% AUTHORS INFORMATION AREA
%***********************************************************************
\author{M.J. Charles$^{1,2}$
% Optional short acknowledgment: remove next line if non-needed
%\thanks{Work supported by DOE.}
%\thanks{This is an optional funding source acknowledgment.}
% DO NOT MODIFY THE FOLLOWING '\vspace' ARGUMENT
\vspace{.3cm}\\
% Addresses and institutions (remove "1- " in case of a single institution)
1- University of Iowa, Iowa City, Iowa 52242, USA \\
2- Oxford University, Oxford, United Kingdom
}
%%***********************************************************************
% END OF AUTHORS INFORMATION AREA
%***********************************************************************

\maketitle

\begin{abstract}
A PFA has been developed for the SiD detector concept
at a future Linear Collider.
The performance of the version of this PFA used in the
SiD LOI is presented for a number of physics processes
with two hadronic jets.
Presented at LCWS08~\cite{url}.
\end{abstract}

\section{Introduction}
\label{sec:pfa:intro}

Reconstruction in SiD is based on the particle flow concept in which calorimeter energy deposits from individual particles are separated, allowing the energy of each to be measured in the optimal subsystem for that particle (the silicon tracker for charged particles, the EM calorimeter for photons, both calorimeters for neutral hadrons). In the limit of perfect separation, the contribution to the jet energy resolution from charged particles is negligible and only neutral hadrons need to have their energy measured in the hadronic calorimeter, leading to a jet energy resolution of roughly\footnote{
  $E$ is in units of GeV throughout.
} $20\%/\sqrt{E}$~\cite{bib:ron}. In practice, this limit is difficult to achieve. Degradation of the resolution due to imperfect separation of energy deposits is generically referred to as confusion, and is the most important effect for well-contained, high-energy jets in the SiD acceptance. A particle flow algorithm (PFA) has been developed and tuned for SiD in the {\tt org.lcsim} software framework with the goal of minimizing the confusion and therefore the resolution. A snapshot of the PFA has been used for the analysis and benchmarking results reported in this LOI; development is still in progress and performance is expected to continue improving in future versions.

A deliberate effort has been made to keep the code as modular as possible. Different components communicate with one another by reading and writing named objects in standard formats to the event-level data store. This makes the flow of information clear, and allows one component to be substituted for another.

For each event, the SiD PFA takes as inputs the energy deposits in the calorimeters and muon system and the set of tracks found in the tracking system (as described in 
Reference~\cite{bib:rich}).
The PFA then performs the reconstruction in a series of steps, 
described in detail in Reference~\cite{bib:taejeong}.
The general strategy for pattern-recognition in the calorimeters is (a) to identify and set aside the easiest, most distinctive showers first, taking maximum advantage of the information, and (b) to recognize common classes of mistakes made earlier in the algorithm and correct for them. The PFA produces as output a collection of reconstructed particles suitable for use in a physics analysis. 

\section{Performance}

\begin{figure}
  \begin{center}
    \includegraphics[width=0.49\columnwidth, trim = 0 1.2cm 1.95cm 1.2cm, clip]{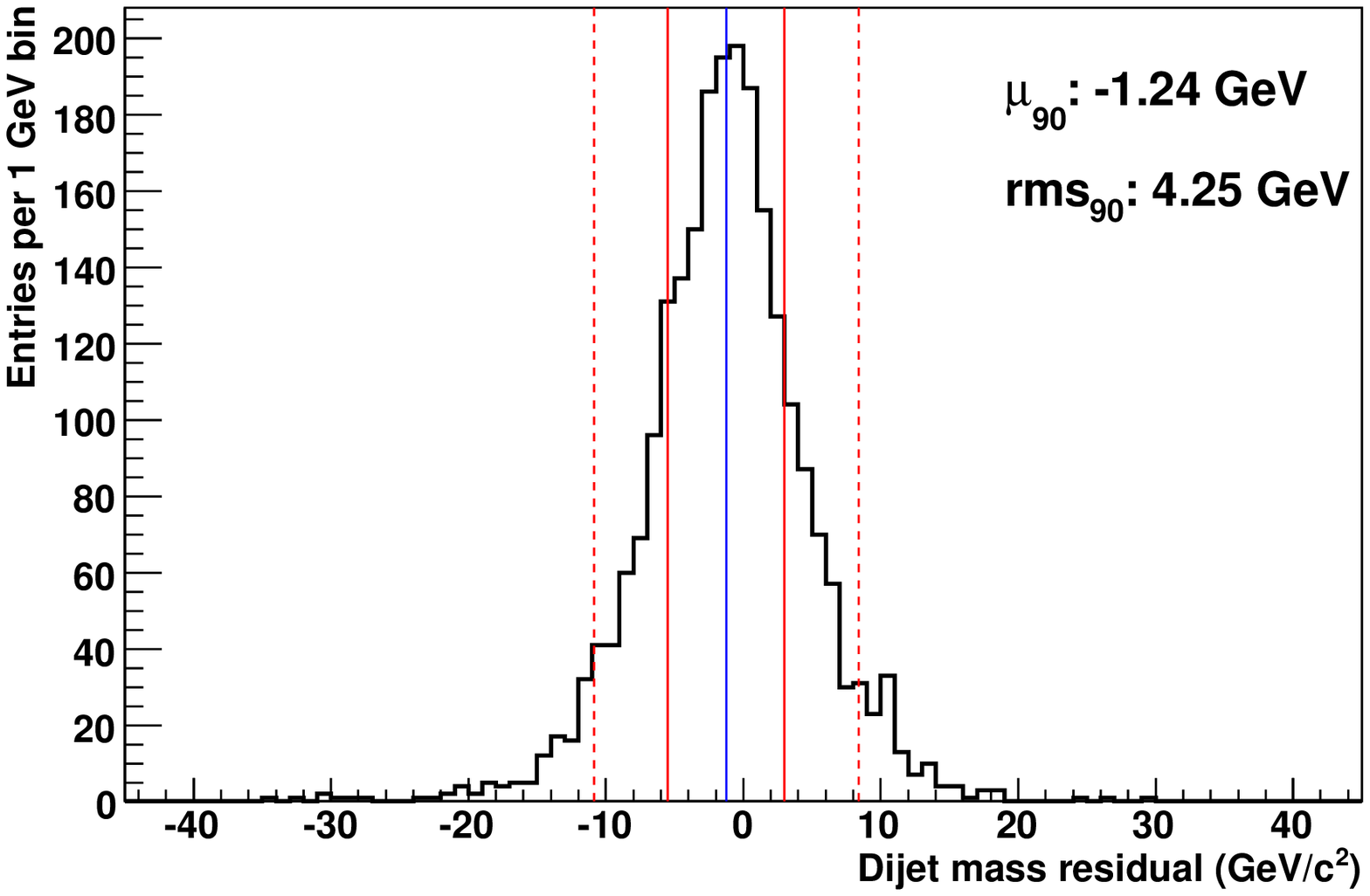}
    \includegraphics[width=0.49\columnwidth, trim = 0 1.2cm 1.95cm 1.2cm, clip]{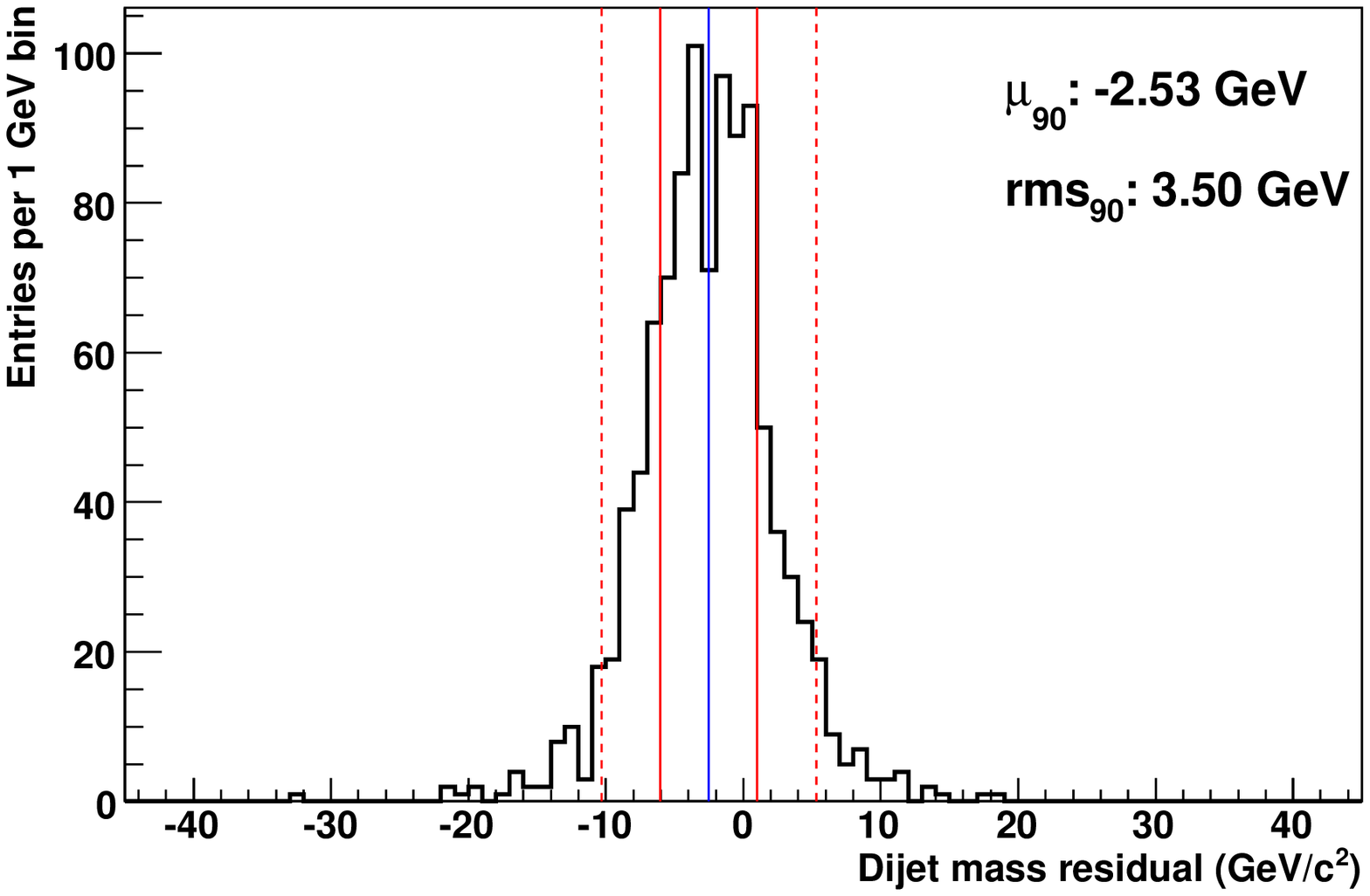}
  \end{center}
\caption{
  Dijet mass residuals for $e^+ e^- \to Z (q \bar{q}) Z (\nu \bar{\nu})$ events at $\sqrt{s}=500$~GeV, shown for the barrel (left) and endcap (right) regions of polar angle.
}
\label{fig:pfa:massplot}
\end{figure}

The true test of performance is the sensitivity to key physics observables---this is discussed in Reference~\cite{bib:tim}. However, for the purposes of studying and optimizing a PFA it is helpful to look at specific physics processes which are simple to analyse and depend primarily on the quality of the PFA output. We use two such processes:
\begin{itemize}
  \item $e^+ e^- \to q \bar{q}$ at $\sqrt{s}=100,200,360,500$~GeV, for $q=u,d,s$. Beamstrahlung and bremstrahlung in the initial state are disabled so that the collision energy $E_{\mathrm{CM}}$ is the same as $\sqrt{s}$. The figure of merit is the event energy sum residual $\Delta E_{\mathrm{CM}}$, i.e. the signed difference between the reconstructed and true values of $E_{\mathrm{CM}}$. Under the simplifying assumption that the invariant mass of two jets with energies $E_1$ and $E_2$ and opening angle $\theta_{12}$ is given by $m_{12}^2 = 2 E_1 E_2 (1 - \cos\theta_{12})$, the resolution of energy sum residuals is equal to the resolution of the dijet mass for jets of the same energy.
  \item $e^+ e^- \to Z (q \bar{q}) Z (\nu \bar{\nu})$ at $\sqrt{s}=500$~GeV, for $q=u,d,s$. The figure of merit is the dijet mass residual $\Delta M$, the signed difference between the reconstructed and true values of $m_{q \bar{q}}$. Plots of the residual distribution are shown in Figure~\ref{fig:pfa:massplot}.
\end{itemize}
In both cases, the figure of merit depends upon the quality of hadronic jet reconstruction but does not require jet-finding or corrections for primary neutrinos.

\begin{wrapfigure}{r}{0.5\columnwidth}
  \centerline{\includegraphics[width=0.5\columnwidth, trim = 0 1.2cm 1.95cm 1.2cm, clip]{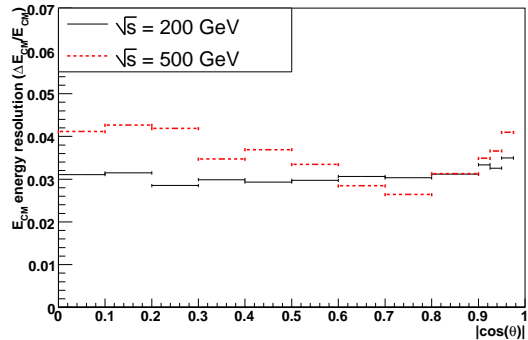}}
  \caption{
    Resolution ($\mathrm{rms}_{90}$ of the energy sum residuals) as a function of angle for $e^+ e^- \to q \bar{q}$ for $\sqrt{s}=200,500$~GeV. For 200~GeV (solid), leakage is not significant and the angular distribution is roughly flat, rising slowly towards $|\cos\theta|=0.975$ as the effects of acceptance and tracking become important. For 500~GeV (dashed), leakage has a major impact---this can be seen from how the resolution varies in the barrel between $|\cos\theta|=0$ where the calorimeter thickness is minimized to $|\cos\theta|=0.8$ where it is greatest. For both energies, the resolution is very bad for $|\cos\theta|>0.975$ due to acceptance losses.
  }
  \label{fig:pfa:angular}
\end{wrapfigure}

Table~\ref{tab:pfa:main-results} shows the measured resolutions\footnote{
 Resolutions are quoted in terms of $\mathrm{rms}_{90}$, the RMS of the contiguous block of 90\% of events with smallest RMS. Similarly,  $\mu_{90}$ is defined to be the mean of these events. Note that for a Gaussian distribution, the $\mathrm{rms}_{90}$ is approximately 78\% of the full RMS.
} for the {\tt sid02} detector. The resolution is quoted separately for the barrel ($0 < |\cos(\theta)| < 0.8$) and endcap ($0.8 < |\cos(\theta)| < 0.95$) regions of polar angle. There are several effects at work:
\begin{itemize}
  \item The calorimetric component of the resolution function is expected to scale as $\sqrt E$, i.e. slower than linear. When this dominates, the fractional resolution ($\sigma_{\Delta E_{\mathrm{CM}}}/E_{\mathrm{CM}}$) decreases as the energy goes up.
  \item The confusion component of the resolution function will increase as the jet energy goes up and pattern-recognition becomes harder. The energy-dependence is not known from first principles, but it is likely to be at least linear if not faster.
  \item At high energies, leakage of energy out of the back of the calorimeter becomes important. The impact on the resolution has a strong angular dependence, since the effective depth of the calorimeter varies with $\cos\theta$; this is illustrated in Figure~\ref{fig:pfa:angular}. This effect is modest for jet energies of 180~GeV but becomes dominant by 250~GeV. It is partially mitigated in the endcap region by using the muon system as a tail-catcher; this depends strongly on the longitudinal segmentation in the muon system and was found to be much more effective with an absorber thickness of 5~cm than 20~cm.
  \item For the $ZZ$ events, the requirement that both jets lie in the angular region of interest constrains the kinematics of the decay.
  \item The dijet mass resolution measured in $e^+ e^- \to ZZ$ events is observed to be larger than the resolution seen in $e^+ e^- \to q\bar{q}$ events, even when the jet energy is comparable. This may be due to non-linearity in the energy response: the $e^+ e^- \to q\bar{q}$ events have mono-energetic jets by construction and so a non-linear response would simply shift the mean of the energy sum distribution, whereas the jets in $e^+ e^- \to ZZ$ events can be quite asymmetric and therefore the dijet mass residual distribution would be broadened by such an effect.
\end{itemize}

\begin{table}
\begin{center}
\begin{tabular}{|l|c|c|c|c|}
 \hline
 & \multicolumn{2}{|c|}{Resolution (real tracking)} & \multicolumn{2}{|c|}{Resolution (cheat tracking)} \\
 \multicolumn{1}{|c|}{Process} 
 & \hspace{0.28cm} Barrel \hspace{0.28cm} 
 & \hspace{0.12cm} Endcap \hspace{0.12cm} 
 & \hspace{0.28cm} Barrel \hspace{0.28cm} 
 & \hspace{0.12cm} Endcap \hspace{0.12cm} 
 \\ \hline
  $e^+ e^- \to q \bar{q}$, $\sqrt{s}=100$~GeV   & 3.7\% & 3.8\% & 3.4\% & 3.5\% \\
  $e^+ e^- \to q \bar{q}$, $\sqrt{s}=200$~GeV   & 3.0\% & 3.2\% & 2.8\% & 3.0\% \\
  $e^+ e^- \to q \bar{q}$, $\sqrt{s}=360$~GeV   & 2.7\% & 2.7\% & 2.6\% & 2.6\% \\
  $e^+ e^- \to q \bar{q}$, $\sqrt{s}=500$~GeV   & 3.5\% & 3.3\% & 3.5\% & 3.4\% \\
  $e^+ e^- \to Z (q \bar{q}) Z (\nu \bar{\nu})$ & 4.7\% & 3.9\% & 4.2\% & 3.7\% \\
  \hline  
\end{tabular}
\end{center}
\caption{
  PFA performance for {\tt sid02}. For the $e^+ e^- \to q \bar{q}$ processes, the $\mathrm{rms}_{90}$ of the energy sum residuals is quoted as a fraction of $\sqrt{s}$, and for the $e^+ e^- \to ZZ$ process the $\mathrm{rms}_{90}$ of the dijet mass residuals is quoted as a fraction of $m_Z$. Resolutions are quoted for the LOI production snapshot and do not include subsequent improvements.
}
\label{tab:pfa:main-results}
\end{table}

The resolutions in Table~\ref{tab:pfa:main-results} are larger than those seen when running the PandoraPFA algorithm on the ILD detector design~\cite{bib:ward}. Understanding this difference is not straightforward: the performance of a PFA and the design of the detector on which it runs are coupled and it is not meaningful to take either in isolation. It is also technically very difficult to run one PFA on the other detector. However, a work-around has been developed: by starting from the LDC00Sc detector and adjusting the calorimeter geometry and layering, we can produce ``SiDish'' detectors which have similar dimensions to {\tt sid02} and run PandoraPFA~v2.01\footnote{
  PandoraPFA has continued to develop in the meantime; the current version is v03-$\beta$.
} on them~\cite{bib:marcel}. Note that the SiDish detectors still use the same detector technology as LDC00Sc, though: a TPC tracker and iron/scintillator HCAL, unlike SiD's silicon tracker and iron/RPC HCAL. 

The $e^+ e^- \to q \bar{q}$ event energy sum resolution in $0.0<|\cos\theta|<0.7$ found when running PandoraPFA on SiDish detectors resembling {\tt sid02} is 3.1\% for $\sqrt{s}=90$~GeV and 2.8\% for $\sqrt{s}=200$~GeV, superior to the performance we find in Table~\ref{tab:pfa:main-results} (3.7\% and 3.0\%, respectively). Part of this difference is due to the difference between {\tt sid02} and the SiDish detectors. In previous studies comparing SiD detectors with scintillator and RPC instrumentation of the HCAL, we found that the scintillator variant had better performance by about 10\% relative (0.3\% absolute). Likewise, the use of a TPC tracker gives more complete information for decays and interactions inside the tracking system (e.g. for $K_S \to \pi^+ \pi^-$); we can place an upper bound on this of 0.3\% for $\sqrt{s}=100$~GeV and 0.2\% for $\sqrt{s}=200$~GeV from studies with cheat tracking. These effects are sufficient to explain most of the observed performance difference between PandoraPFA and the SiD PFA.

\section{Conclusions}

There has been a great deal of progress in SiD reconstruction since ALCPG07. We have switched to full track reconstruction and found that PFA performance in $e^+ e^- \to q \bar{q}$ events remains close to that of cheat tracking. The PFA itself has been largely rewritten and gives event energy sum resolutions of order 3.0--3.5\% for jet energies up to 250~GeV. The PFA performance was found to be approaching that of the gold standard, PandoraPFA, when running on a comparable detector design for jet energies up to 200~GeV. This is very encouraging for the jet physics prospects at SiD.

Nonetheless, there is a great deal of improvement still to come. A number of code fixes have already been made~\cite{bib:taejeong} and more substantial revisions such as the integration of calorimeter-assisted tracking~\cite{bib:dmitry} are planned. At the broadest level, the two principal challenges are: (1) to understand the impact of leakage in high-energy jets on the physics potential of the detector, and to reduce it by adapting the algorithm and detector design if needed; and (2) to improve the reconstruction algorithm, and in particular to reduce the dijet mass resolution seen in $e^+ e^- \to ZZ$ events. The confusion term still dominates the resolution for the range of jet energies likely to be used in physics analyses at a 0.5~TeV or even 1~TeV collider: we have plenty of room for improvement.

%%%%%%%%%%%%%%%%%%

%\section{Acknowledgments}
%
%To acknowledge funding bodies etc., a special section may be placed
%before the bibliography: \verb?\section*{Acknowledgements}?.

% ****************************************************************************
% BIBLIOGRAPHY AREA
% ****************************************************************************

\begin{footnotesize}

\end{footnotesize}

% ****************************************************************************
% END OF BIBLIOGRAPHY AREA
% ****************************************************************************

\end{document}